\documentclass[12pt]{article}
\usepackage{amsmath}
\usepackage{amssymb}
\usepackage{graphicx}
\usepackage{epsfig}
\newcommand{\bi}[1]{\bibitem{#1}}
\newcommand{\sql}{\sqrt{|\Lambda|}}
\newcommand{\abl}{|\Lambda|}

\newcommand{\be}{\begin{eqnarray}}
\newcommand{\ee}{\end{eqnarray}}
\newcommand{\ra}{\rightarrow}
\newcommand{\e}{\epsilon}

\begin{document}
\baselineskip=14.5pt
\pagestyle{plain}
\begin{titlepage}

\begin{flushright}
PUPT-1968\\
hep-th/0011177
\end{flushright}
\vfil

\begin{center}
{\huge The Emergence of Localized Gravity}\\ [8pt]
\end{center}

\vskip 0.4cm
\begin{center}
{\large Matthew D. Schwartz$^*$}\\
{\it Joseph Henry Laboratories \\
 Princeton University, Princeton, NJ 08544}\\
\end{center}

\vskip 1.2cm
\begin{center}
{\large Abstract}
\end{center}
\noindent
We explore physics on the boundary of a Randall-Sundrum type model when
the brane tension is slightly sub-critical. We calculate the masses
of the Kaluza-Klein decomposition of the graviton and use a toy
model to show how localized gravity emerges as the brane tension becomes
critical. Finally, we discuss some aspects of the boundary conformal field
theory and the AdS/CFT correspondence.

\vskip 0.4cm
{\vskip 5pt \footnoterule\noindent
{\footnotesize $\,^*$\ {\tt
matthew@feynman.princeton.edu}}}
\end{titlepage}
\newpage

\section{Introduction}
There have been a number of proposals to show how four-dimensional gravity might appear in
a universe with five non-compact dimensions. Such models hope to resolve 
fundamental problems in particle physics and cosmology, such as the
hierarchy problem or the cosmological constant problem. For example, Randall and 
Sundrum \cite{rs} demonstrated that a 3-brane embedded in five-dimensional
anti-deSitter space, with fine tuned brane tension supports a single normalizable
massless bound state. Others have shown that if certain other assumptions are made,
there is quasilocalization: gravity is four-dimensional only at intermediate distances\cite{grs}.
The 4D potential in this case comes from an effective resonance of the massive graviton
modes\cite{csaki}. But the quasilocalized models all involve unphysical matter, such as a brane of negative
tension. Recently, Karch and Randall showed that four-dimensional gravity is possible
in a physical scenario which is not so fine tuned \cite{kr}. They allow for a brane with a slightly sub-critical
tension in an $AdS_5$ background. A compact version of their scenario with similar phenomenolgy
has been investigated by Kogan et al. \cite{kogan}\\

In this paper, we explore in more detail the boundary physics of the Karch-Randall model.
The induced metric on the sub-critical brane will have a non-vanishing 4D cosmological constant
$\Lambda < 0$, so it describes $AdS_4$. We will be mostly interested in the 
realistic regime $\abl \ll 1$ (in Planck units) which is not completely incompatible with
observations. In this limit, the brane supports a massive, normalizable
bound state, whose mass is of order $\abl$. Thus, it is effectively massless at physical distances.
There are also a discrete set of massless modes, whose amplitude on the brane are highly suppressed
compared to the bound state.\\ 

\section{Locally Localized Gravity}
First, we briefly review the set-up and results of \cite{kr}.
We start with anti-deSitter spacetime with cosmological constant $\Lambda_{5d} = -3/L^2$. We
put a brane of positive tension $\lambda = \frac{3}{L}\sqrt{1-L^2\abl}$ at $r=0$, and
impose orbifold boundary conditions. We can then explore the transition from five dimensional
anti-deSitter gravity to 4D localized gravity of the Randall-Sundrum model. 
The transverse traceless modes of the graviton satisfy:
\be
\left(\partial_r^2 + 4A'\partial_r -e^{-2A}(\square_{4d}-2\abl)\right)h = 0
\ee
where $A(r)$ is the AdS warp factor. The metric is:
\be
ds^2 = e^{2A}dx^2-dr^2 = e^{2A}(dx^2 -dz^2) \label{met}
\ee
The second form is in conformal coordinates, defined so that $dz^2 = e^{-2A}dr^2$.
The graviton splits up into eigenstates of the 4D d'Alembertian. That is:
\be
(-\square_{AdS4}-2\abl)h_m = m^2 h_m
\ee
In conformal coordinates, the warp factor is:
\be
e^{A(z)} = \frac{L\sql}{\sin\left(\sql(|z|+z_0)\right)}, \quad \sin(\sql z_0) = \sql L
\ee
Still following \cite{kr}, we make the substitution $\psi_m(z) = e^{\frac{3}{2}A(z)} h_m$ and get:
\be
(-\partial_z^2 + V(z) - m^2) \psi_m = 0
\ee
where
\be
V(z) &=& \frac{9}{4} A'(z)^2 + \frac{3}{2} A'{}'(z) \\
&=& -\frac{9}{4}\abl + \frac{15}{4}\frac{\abl}{\sin^2\left(\sql (|z| + z_0)\right)}
-3\sql\cot(\sql z_0)\delta(z) \nonumber
\ee
Finally, we rescale $z$ to $w = \sql z$ and define $\e \equiv \sql z_0 \approx \sql L$.
Then,
\be
(-\partial_w^2 + V(w))\psi_m = E\psi_m, \quad V(w) = -\frac{9}{4} + \frac{15}{4} \frac{1}{\sin^2(|w|+\e)}
-3\cot(\e)\delta(w) \label{exactse}
\ee
$E$ is related to the graviton masses as $m^2 = E\abl$.
If we transfer the AdS bound on $r$ to these coordinates, we see that $0 < w < \pi - \e$. For zero
tension (pure $AdS_5$), $\e=\pi/2$, and as the tension becomes critical, $\e\ra0$.\\

Equation \eqref{exactse} determines the form of the graviton excitations. It has the form
of the non-relativistic Schroedinger equation, and we can solve it exactly to show that each graviton
mode is some linear combination of two hypergeometric functions. But then it is quite a messy task
to impose the delicate boundary conditions. Rather than grind out the analytic solution, if there
is one, we can try to understand the modes by looking at a similar potential in an exactly solvable toy model.
We can then verify that it has many of the same qualitative features, and gain some intuition
for how localization comes about.

\subsection{Toy model}
The idea is to approximate the potential in \eqref{exactse} with $0 < w < \pi - \e$ and $\e \ll 1$. 
Note that the potential blows up at the AdS boundary, $w = \pi-\e$, but is finite, although very
large, at the Planck brane $w=0$. It also has a large (but finite) delta function on the brane.
So, we choose our toy potential to contain a wall at $w=\pi-\e$ and a delta function at 
$w=0$ of the right strength: $-3\cot(\e)\delta(w)$. And we will impose orbifold symmetry on the wavefunctions,
to match the real situation. In other words, to get the toy potential, we set the real potential to zero wherever it is finite.\\

The exact solutions to this toy model are a set of massive sinusoidal modes, which vanish
at the AdS boundary.
\be
\chi_k(w) = \sin\left(k(\pi - \e - w)\right)
\ee
The energies are $E = k^2$ where $k$ satisfies:
\be
\frac{2}{3}\tan\e = \frac{\tan(k(\pi-\e))}{k}
\ee
For the pure AdS case ($\e = \pi/2$) there are only odd modes, $k = 1,3,5,\cdots$. As $\e$ goes
to zero, ($\abl \ra 0$), the frequencies decrease, and we end up with all the integers, $k = 1,2,3,\cdots$.\\

\begin{figure}
\begin{center}
\includegraphics{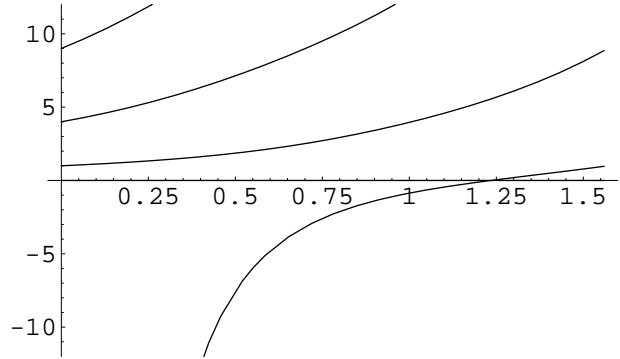}
\caption{The energies $E$ of the first few eigenfunctions of the toy potential as a function of $\e$. The emergence
of the negative energy bound state can be seen as the brane tension is increased ($\e \ra 0$).}
\label{toyenergies}
\end{center}
\end{figure}

In addition, once $\e$ is below the critical value
\be
(\frac{3}{2}\cot\e )(\pi-\e) = 1 \Rightarrow \e = 1.2345 \label{ecrit} 
\ee
the delta function supports a bound state with energy $E = -\kappa^2$. It has the form:
\be
\chi_0(w) =  \sinh\left(\kappa(\pi - \e - w)\right) \approx e^{-\kappa w}
\ee
where $\kappa$ satisfies:
\be
\frac{2}{3}\tan\e = \frac{\tanh(\kappa(\pi-\e))}{\kappa} \Rightarrow \kappa \approx \frac{3}{2}\cot\e
\approx \frac{3}{2\e}
\ee
The energies $E$ of the first few modes for the toy model are shown in figure \eqref{toyenergies}.\\

\begin{figure}
\begin{center}
\includegraphics{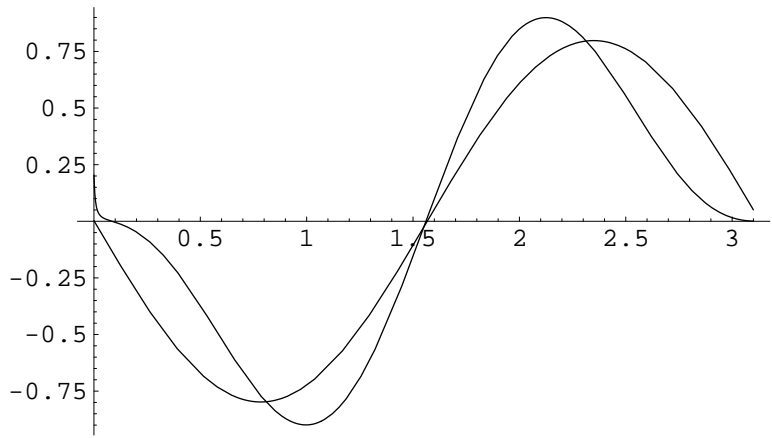}
\caption{Comparison of the toy model and exact wavefunction (solved numerically), for n=2 and $\e = 0.01$. The value
of the normalized wavefunctions are shown as a function of $w$. The brane is at $w=0$ and the boundary of $AdS_5$ at
$w=\pi-\e$.}
\label{toyreal}
\end{center}
\end{figure}

We can compare these results to numerical solutions of \eqref{exactse} (see figure \eqref{toyreal}). The massive modes
are qualitatively very similar, except near $w=0$. The exact masses go like $n(n+3)$ (see below), so the 
toy model correctly predicts this quadratic behavior.  Finally, there is also a bound state for small $\e$ which is exponentially
bound to the brane, just like in the toy.\\

What does this toy model tell us? Recall that it is the bound state which is responsible for localized
4D gravity. This follows because a bound state will have a much higher amplitude on the brane than
an unbound state, after normalization. Thus, if we understand how the bound state arises in the toy
model, we will understand how localized gravity appears as we increase the brane tension.
If we start with a zero tension brane, the lightest mode ($k=1$) looks like the first
quarter period of $\cos(w)$. The AdS boundary condition pins the right side to zero. As the brane tension
is increased, the wavefunction must maintain $\chi'(0) = -\frac{3}{2}\cot(\e)\chi(0)$
because of the $\delta$-function. This causes the cos curve to flatten ($k\ra0$). Eventually,
it can't flatten anymore and inverts ($\sin\ra\sinh$), leading to the bound state. The same
qualitative transformation occurs in the real (not toy) case, and gives us some intuition
for why 4D gravity emerges.\\

The toy model does have flaws.
For example, the it predicts the mass of the
bound state should head towards $-\infty$ in the critical limit, while we know that the
real bound state has a positive mass (this is gauranteed by a simple argument from supersymmetry,
see for example \cite{mim}) and becomes massless in the critical limit.
We will return to this below.
There are also a number of important questions about the real system which the toy model
cannot answer.
For example, we want to calculate the effective potential on the brane. To do this by summation,
we need to know the amplitude of the graviton modes on the Planck brane. But the Planck brane is precisely
where the discrepancy between the toy potential and the real potential is largest. We would also
like to know the mass of the bound state, but since it is localized on the brane, having a good
representation of the potential there is critical. Therefore, we must look at the solutions to
the original problem. However, as mentioned above, and in \cite{kr}, these solutions are a set
of ugly hypergeometric functions which refuse even to reveal their small $\e$ limit. Instead,
we can attempt understand the solutions numerically.\\

\subsection{Numerical results}
\begin{figure}
\begin{center}
\includegraphics{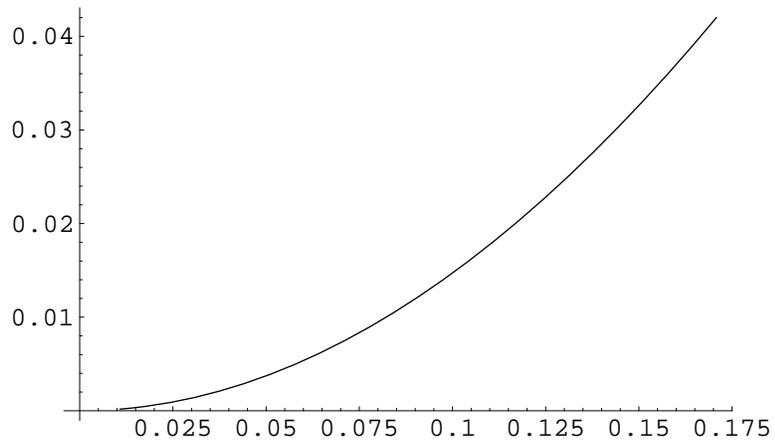}
\caption{Mass of the bound state ($E_0$) as a function of $\e$, calculated numerically.}
\label{masscurve}
\end{center}
\end{figure}

\begin{figure}
\begin{center}
\includegraphics{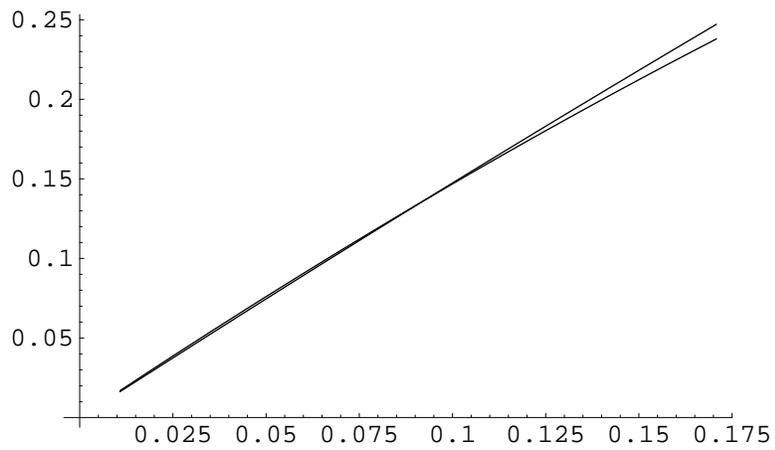}
\caption{$E_0/\sin(\e)$ (top) compared to $(1.5 - \e)\sin\e$ as a function of $\e$.}
\label{massstraight}
\end{center}
\end{figure}

By applying the shooting method to the analog quantum mechanics problem, we can estimate the
dependence of the bound state mass on the brane tension. This is actually quite a delicate
calculation since as the tension becomes critical ($\e\ra 0$), the potential becomes singular. Instead
of taking $\e$ as small as possible, we plot the mass as a function of $\e$ and extrapolate 
(fig \eqref{masscurve}). It seems to be a very good approximation (fig \eqref{massstraight})
to say that:
\be
E_0 &\approx& (1.5 - \e)\sin^2\e \ra 1.5 \e^2\\
\Rightarrow m_0^2 &\approx& 1.5\abl^2
\ee
This result can also be obtained from an analytic argument invovling a supersymmetric decomposition
of the volcano potential \cite{mim}.

The excited states have non-vanishing zeroth order (in $\e$) terms. The
first order terms seem to vanish and the spectrum looks like:
\be
E_n &\approx& n(n+3) + 0.4 n^3 \e^2 + {\cal O}(\e^3)\\
\Rightarrow m_n^2 &\approx& n(n+3)\abl +0.4 n^3 \abl^2+ {\cal O}(\abl^3), \:n = 1,2,3,\cdots 
\ee

To estimate the amplitude of the graviton modes on the Planck brane, we need know their normalizations.
These are determined by the requirement that:
\be
1 = \int \sqrt{-g} h^2dz = \int\psi^2 dz \Rightarrow \int \psi^2 dw = \sql
\ee
where we used AdS metric \eqref{met}.  Using this normalization, we can numerically estimate the
amplitudes of the wavefunction at the brane:
\be
\psi_0^2(0) &\approx& 2-0.1\e + {\cal O}\left(\e^2\right)\\
\psi_n^2(0) &\approx& 2.5 \e^2 + 0.8 (n\e^2) + {\cal O}\left(\e^3,n^2\right)
\ee
This lets us approximate the gravitational potential on the
brane. Because the amplitudes of the excited modes are much smaller than the near zero
mode, they will just give order $\abl^2$ corrections to anti-deSitter gravity, which 
itself is well approximated by Minkowski gravity, up to distances of order $\frac{1}{\abl}$. In
such a regime, it is reasonable to approximate the potential generated by an AdS graviton
as Yukawa, which makes the total effective potential have the form:
\be
V(r) = \psi_0^2(0) \frac{e^{-m_0 r}}{r} + \sum_n \psi_n^2(0)\frac{e^{-m_n r}}{r}
\ee

As $\abl \ra 0$, the masses approach a continuum, and we can replace the sum with an integral.
Then, dropping constants of order unity,
\be
V(r) &\approx& \frac{1}{r}e^{-\abl r} + \frac{\abl}{r}\int (1+n) e^{-n\sql r}dn\\
 &\approx& \frac{1+\sql}{r} + \frac{\sql}{r^2} + \frac{1+\sql}{r^3} + \cdots
\ee
In the limit that $\Lambda = 0$, we get precisely the corrections RS found for the critical case. These numerical
results just show that each term in the potential (including the $0/r^2$ term) gets a correction of order $\sql$.

\section{AdS/CFT}
It is well established by now that gravitational theories in anti-deSitter space are dual to conformal
field theories on the boundary \cite{adscft}. In the Randall-Sundrum model, the presence of a brane at $r=0$ changes
the 5D gravity from pure $AdS$, especially at small $r$. Correspondingly, since small $r$ corresponds to
high energy in the CFT, there is an induced cutoff in the UV. This breaks scale invariance and induces
gravity (see, for example, \cite{nunez}). However, the gravity is weak, and at low energies, the theory is very nearly conformal. Following
a suggestion by Witten, some authors have calculated the corrections to 4D gravity induced by the CFT
and shown that it precisely matches the corrections from the high energy 5D graviton modes\cite{gub,duff}.
In fact, the calculation is greatly simplified because the graviton propagator is corrected only by the $TT$ correlator, which
is determined completely by conformal invariance.\\

So, it is natural to ask if there is a dual description of the boundary theory in the current case.
As Karch and Randall have shown, the brane only cuts off part of the boundary. So the boundary of
the 5D theory consists of (A) the brane and (B) half of the normal boundary.  
(A) is a complete anti-deSitter space with a small negative cosmological constant of magnitude $\abl$, while
(B) is Minkowski space artificially cut off at some finite distance.  The dual conformal field theory must take place
on this background. It is known that the residual symmetry on (B) is an $SO(3,2)$ subgroup of the full
conformal group\cite{mcavity}, so the symmetry groups on the two sections are the same.\\

The boundary conformal field theory gives us a very interesting perspective on the
$\abl\ra 0$ limit. We know that as we reduce $\abl$ (by increasing the brane tension), 
the brane moves closer to the $AdS_5$ boundary (see section 2).
So if we actually take this limit, it seems that part (A) becomes
the other half of the boundary and we just end up with the whole thing: Minkowski space. This
corresponds to a particular value of $r$ for the position of the Minkowski brane in the RS case:
at the $AdS_5$ boundary ($r=\infty$). If we take this limit in the CFT dual of Randall-Sundrum, we see that
the UV cutoff, which corresponds to the position of the brane, goes to infinity. In either case,
the full conformal symmetry is restored: either we remove the spatial cutoff or the energy cutoff.
Therefore, the only way to go between the RS CFT dual and KR CFT dual is by restoring the conformal
symmetry first. We can understand why this must be true from
the group theory side: the Poincare group, $ISO(3,1)$ is a subgroup of $SO(4,2)$.
The anti-deSitter group, preserved on the sub-critical brane is an $SO(3,2)$ subgroup. We can
contract this group to get the Poincare group, but it will not be the same $ISO(3,1)$ as the
one preserved on the Minkowski slice. After all, one is a subgroup of $SO(4,2)$ and the other is not. The
only way we can get them to coincide is if we restore the full $SO(4,2)$ symmetry first, and
then take the subgroup in another direction.\\

It is also interesting to look at how a CFT might correct $AdS_4$ gravity.
Suppose we are living on the brane part (A), and that there are a set of massive gravitons with
masses $m_0^2 = n(n-3)\abl$. The conformal field theory would induce corrections to these masses
proportional to $\abl^2$. To see this, notice that the $<TT>$ correlation function goes like
\be
M^2(p^2) = \langle T T \rangle = c p^4 \log p
\ee
where $c$ is the central charge. By integrating out the 1-loop diagrams, this
produces a correction to the mass as: 
\be
\delta m^2 = M^2(m_0^2) \propto m_0^4 \propto \abl^2
\ee
and thus gravity in such a scenario would be very similar to induced gravity on the brane,
which we described in section 2.
Now we must ask where the original ${\cal O}(\abl)$ masses come from. One possibility
is that they are induced by the CFT on (B), although we cannot see any obvious reason for this
effect. If the limit where the boundary of the 4D space goes to infinity and the full conformal
symmetry is restored, the massless graviton would have to transform in a representation of the
full conformal group, not its $SO(3,2)$ subgroup. As shown in \cite{kr} the $(4,1,1)$ representation of
$SO(4,2)$ decomposes into a set of irreducible $SO(3,2)$ representations with the spectrum
$m^2 = E(E-3)\abl$. So, perhaps the ${\cal O}(\abl)$ terms come from the breaking of conformal invariance
and the higher order corrections come from the residual conformal field theory. This is certainly
an interesting area for future work.\\

\section{Acknowledements}
I would like to thank Andreas Karch and Lisa Randall for collaborating on most of this work. In addition, I
thank S. Gubser for helpful discussions.

\end{document}